\documentclass[aps,prd,twocolumn,showpacs,amsmath,amssymb]{revtex4-1}
\usepackage{amsmath}
\usepackage{graphicx}
\usepackage{subfigure}
\usepackage{epstopdf}
\usepackage{color}
\usepackage{multirow}
\usepackage{setspace}
\usepackage{overpic}
\usepackage{amssymb}
\usepackage{subfigure}
\usepackage[bookmarksnumbered, pdfstartview=FitH,colorlinks,urlcolor=blue, citecolor=blue,linkcolor=blue] {hyperref}
\usepackage{lineno}
\usepackage{bm}
\usepackage{rotating}
\usepackage[utf8]{inputenc}
\usepackage{cancel}
\let\oldequation\equation
\let\oldendequation\endequation

\renewenvironment{equation}
  {\linenomathNonumbers\oldequation}
  {\oldendequation\endlinenomath}

\begin{document}


\title{\bf Observation of $D^+\to f_0(500)\mu^+\nu_\mu$ and study of $D^+\to \pi^+\pi^-\ell^+\nu_\ell$ decay dynamics}

\author{M.~Ablikim$^{1}$, M.~N.~Achasov$^{4,b}$, P.~Adlarson$^{75}$, O.~Afedulidis$^{3}$, X.~C.~Ai$^{80}$, R.~Aliberti$^{35}$, A.~Amoroso$^{74A,74C}$, Q.~An$^{71,58}$, Y.~Bai$^{57}$, O.~Bakina$^{36}$
% ...
\\
\vspace{0.2cm}
(BESIII Collaboration)\\
\vspace{0.2cm} {\it
$^{1}$ Institute of High Energy Physics, Beijing 100049, People's Republic of China\\
% ...
$^{3}$ Bochum  Ruhr-University, D-44780 Bochum, Germany\\
$^{4}$ Budker Institute of Nuclear Physics SB RAS (BINP), Novosibirsk 630090, Russia\\
% ...
$^{35}$ Johannes Gutenberg University of Mainz, Johann-Joachim-Becher-Weg 45, D-55099 Mainz, Germany\\
$^{36}$ Joint Institute for Nuclear Research, 141980 Dubna, Moscow region, Russia\\
% ...
$^{57}$ Southeast University, Nanjing 211100, People's Republic of China\\
$^{58}$ State Key Laboratory of Particle Detection and Electronics, Beijing 100049, Hefei 230026, People's Republic of China\\
% ...
$^{71}$ University of Science and Technology of China, Hefei 230026, People's Republic of China\\
% ...
$^{74}$ University of Turin and INFN, (A)University of Turin, I-10125, Turin, Italy; (B)University of Eastern Piedmont, I-15121, Alessandria, Italy; (C)INFN, I-10125, Turin, Italy\\
$^{75}$ Uppsala University, Box 516, SE-75120 Uppsala, Sweden\\
% ...
$^{80}$ Zhengzhou University, Zhengzhou 450001, People's Republic of China\\
\vspace{0.2cm}
% ...
$^{b}$ Also at the Novosibirsk State University, Novosibirsk, 630090, Russia\\
% ...
}
}

\begin{abstract}
Using $2.93~\mathrm{fb}^{-1}$  of $e^+e^-$ collision data collected with the BESIII detector at the center-of-mass energy of 3.773 GeV,
we investigate the semileptonic decays $D^+\to \pi^+\pi^- \ell^+\nu_\ell$ ($\ell=e$ and $\mu$).
The $D^+\to f_0(500)\mu^+\nu_\mu$ decay is observed for the first time.
By analyzing simultaneously the differential decay rates of
$D^+\to f_0(500) \mu^+\nu_\mu$ and $D^+\to f_0(500) e^+\nu_e$ in different $\ell^+\nu_\ell$ four-momentum transfer intervals,
the product of the relevant hadronic form factor $f^{f_0}_{+}(0)$ and the magnitude of the $c\to d$ Cabibbo-Kobayashi-Maskawa matrix element $|V_{cd}|$
is determined to be $f_{+}^{f_0} (0)|V_{cd}|=0.143\pm0.014_{\rm stat}\pm0.011_{\rm syst}$ for the first time.
With the input of $|V_{cd}|$ from the global fit in the standard model, we determine $f_{+}^{f_0} (0)=0.63\pm0.06_{\rm stat}\pm0.05_{\rm syst}$.
The absolute branching fractions of $D^+\to f_0(500)_{(\pi^+\pi^-)}\mu^+\nu_\mu$ and $D^+\to \rho^0_{(\pi^+\pi^-)} \mu^+\nu_\mu$ are determined as
$(0.72\pm0.13_{\rm stat}\pm0.08_{\rm syst})\times10^{-3}$ and $(1.64\pm0.13_{\rm stat}\pm0.10_{\rm syst})\times 10^{-3}$.
Combining these results with those of previous BESIII measurements on their semielectronic counterparts from the same data sample,
we test  lepton flavor universality by measuring the branching fraction ratios
${\mathcal B}_{D^+\to \rho^0 \mu^+\nu_\mu}/{\mathcal B}_{D^+\to \rho^0 e^+\nu_e}=0.88\pm0.10$ and
${\mathcal B}_{D^+\to f_0(500) \mu^+\nu_\mu}/{\mathcal B}_{D^+\to f_0(500) e^+\nu_e}=1.14\pm0.26$,
which are compatible with the standard model expectation.

\end{abstract}

\maketitle

\oddsidemargin  -0.2cm
\evensidemargin -0.2cm

\section{Motivation}

A series of ground breaking discoveries of the $XYZ$ states~\cite{pdg2022}, implying tetraquark composition, at the turn of this century
 and the subsequent observation of pentaquarks $P_c$~\cite{Pc2015,Pc2019} have revolutionized hadron spectroscopy. These related topics have been the focus of much experimental and theoretical interest in recent years.
 Quantum Chromodynamics (QCD) has been established for half a century as the fundamental theory of the strong interaction,
whose development has benefited from such spectroscopy studies since the beginning.
However, it is surprising that the nature of the light scalar mesons, which constitute the spectrum of lowest mass states,  is still under debate after almost seven decades~\cite{pdg2022,Pelaez2016,DLYao2021}.
Notably, they are located near the $S$-wave thresholds of two-body  final states, such as $KK$ and $\pi\pi$.
The ongoing puzzle behind these phenomena can be ascribed to the nonperturbative dynamics at low energies and quark confinement.

The lightest scalar meson $f_0(500)$, traditionally known as the $\sigma$ meson, was not well-established for many years,
with consensus on its mass and width only being reached in the past decade or so~\cite{Pelaez2016}.
The list of references is huge and more information can be found in the Review of Particle Properties~\cite{pdg2022},
and recent reviews~\cite{Pelaez2016,Pelaez2022}.
Sharing the same spin-parity quantum numbers with the QCD vacuum,
this meson plays a significant role in the dynamical generation of mass through the spontaneous breaking of QCD chiral symmetry, and consequently the confinement of quarks.
Clarifying the nature of the $f_0(500)$ could shed light on these issues.
However, a $q\bar{q}$ configuration of the $f_0(500)$ and other light scalar mesons forming the SU(3) nonet in the naive quark model cannot explain their inverted mass hierarchy.
There is still the possibility that these are mixtures of $q\bar{q}$ states~\cite{sigma-mix}.
Other interpretations include diquark-antidiquark states~\cite{sigma-tetraquark},
meson-meson bound states~\cite{sigma-molecular}, and even more complicated scenarios~\cite{sigma-complicated}.

The form factor (FF) of semileptonic (SL) $D^+$ transition to the $f_0(500)$ not only provides new information about the nature of the $f_0(500)$
but also helps to understand the dynamics of SL charmed-meson decays by testing different nonperturbative theoretical methods.
Predictions exist of the branching fractions (BFs)~\cite{RMWang2023,f0_bf,YKHsiao2023}
and the FF~\cite{Dosch:2002rh,Gatto2000,chPT} of $D^+\to f_0(500)\ell^+\nu_\ell$ ($\ell=e$ and $\mu$).
Additionally, scrutinizing the BF ratios of SL $D$ decays allows for probes of lepton flavor universality (LFU), and therefore tests the standard model (SM) in the charm sector~\cite{KBCLS}.

To date, the $D^+\to f_0(500)\mu^+ \nu_\mu$ decay has not yet been observed, and
only the FOCUS, E791, and E687 experiments have performed  studies of $D^+\to \rho^0 \mu^+\nu_\mu$~\cite{FOCUS,E791,E687};
 while the BESIII experiment has observed $D^+\to f_0(500) e^+\nu_e$~\cite{bes3-rhoev}.
This paper reports the observation and BF measurement of $D^+\to f_0(500)\mu^+ \nu_{\mu}$, studies of  the dynamics  of
$D^+\to f_0(500)\ell^+ \nu_\ell$, the absolute measurement of the BF of $D^+\to \rho^0 \mu^+ \nu_{\mu}$,
and comprehensive LFU tests with $D^+\to \rho^0\ell^+ \nu_\ell$ and $D^+\to f_0(500)\ell^+ \nu_\ell$, for the first time.
These analyses make use of 2.93~fb$^{-1}$ of $e^+e^-$ collision data taken at the center-of-mass energy of 3.773 GeV.
Throughout this paper, charge conjugate channels are always implied.
Due to slightly different event selection criteria with Ref.~\cite{bes3-rhoev} as mentioned later,
the BFs of $D^+\to \pi^+\pi^-e^+\nu_e$, $D^+\to f_0(500) e^+\nu_e$, and $D^+\to f_0(500) e^+\nu_e$ obtained
in this work offer independent checks; while the hadronic FF of $D^+\to f_0(500) \ell^+\nu_\ell$ is measured for the first time.

\section{BESIII detector and Monte Carlo simulations}

The BESIII~\cite{BESCol} detector is a magnetic spectrometer located at the Beijing Electron Positron Collider
(BEPCII)~\cite{BEPCII}. More details about the BESIII detector are described in Ref.~\cite{BESCol}.
Monte Carlo (MC) simulated data samples, produced with a {\sc geant4}-based~\cite{geant4} software package including the geometric description of the BESIII detector~\cite{detvis} and the
detector response, are used to determine the detection efficiencies
and to estimate the background contributions. The simulation includes the beam-energy spread and initial-state radiation in the $e^+e^-$
annihilations modeled with the generator {\sc kkmc}~\cite{kkmc}.
The inclusive MC sample consists of the production of $D\bar{D}$
pairs with consideration of quantum coherence for all neutral $D$
modes, the non-$D\bar{D}$ decays of the $\psi(3770)$, the initial-state radiation
production of the $J/\psi$ and $\psi(3686)$ states, and the
continuum processes.
The known decay modes are modeled with {\sc
evtgen}~\cite{evtgen} using the BFs taken from the
PDG~\cite{pdg2022}, and the remaining unknown decays
from the charmonium states are modeled with {\sc
lundcharm}~\cite{lundcharm}. Final-state radiation
from charged final state particles is incorporated with the {\sc
photos} package~\cite{photos}.
The SL decays $D^+\to \pi^+\pi^-\ell^+\nu_\ell$ are simulated according to the results of a  previous BESIII amplitude analysis measurement in the $e$ channel~\cite{bes3-rhoev}.

\section{Single-tag $\bar D$ candidates}
A detailed description of the selection criteria for charged and neutral particles is provided in Refs.~\cite{epjc76,cpc40,bes3-pimuv}.
The tagged $D^-$ mesons are reconstructed in six hadronic final states
$K^{+}\pi^{-}\pi^{-}$, $K^{+}\pi^{-}\pi^{-} \pi^{0}$, $K_{S}^{0}\pi^{-}$, $K_{S}^{0}\pi^{-}\pi^{0}$, $K_{S}^{0}\pi^{-}\pi^{+}\pi^{-}$ and $K^+K^-\pi^-$.
The $D^-$ signals are separated from backgrounds with two kinematic variables: $\Delta E \equiv E_{D^-} - E_{\rm beam}$ and
$M_{\rm BC} \equiv \sqrt{E_{\rm beam}^2 -  |\vec p_{D^-}|^2}$,
where $E_{D^-}$ and $\vec p_{D^-}$ are the reconstructed energy and momentum of the $D^-$ candidate in the $e^+e^-$ center-of-mass frame, and $E_{\rm beam}$ is the beam energy.
If there are multiple candidates per tag mode, the one with the minimum $|\Delta E|$ is kept.
For each tag mode, the yield is extracted by fitting the $M_{\rm BC}$ distribution as in Refs.~\cite{epjc76,cpc40,bes3-pimuv}.
The total yield of tagged $D^-$ mesons is $N^{\rm tot}_{\rm tag}=(1522.5\pm 2.2_{\rm stat.})\times 10^3$.

\section{Double-tag events}

After a tagged candidate is selected, the $D^+\to \pi^+\pi^- \ell^+ \nu_\ell$ decays are reconstructed recoiling against the tagged $D^-$.
This requires a $\ell^+$ candidate and a $\pi^+\pi^-$ pair in the signal side.
The $\pi^\pm$ and $e^+$ are identified with the same criteria as Ref.~\cite{epjc76}.
Particle identification (PID) for the $\mu^+$ considers the measurements of the specific ionization energy loss by the main drift chamber (MDC),
the flight time by the time-of-flight system, and the energy deposited in the electromagnetic
calorimeter~(EMC). Based on these measurements, we calculate the combined confidence levels for positron ($CL_e$), muon ($CL_\mu$), pion ($CL_\pi$), and kaon ($CL_K$) hypotheses for each charged track. The charged track satisfying $CL_{\mu}>0.001$,  $CL_{\mu}>CL_e$, $CL_{\mu}>CL_\pi$, $CL_{\mu}>CL_K$,
and $E_{\rm EMC} \in (0.09, 0.31)~\rm GeV$ is assigned as $\mu^+$ candidate,
where $E_{\rm EMC}$ is its energy deposited in the EMC.
The two requirements $CL_{\mu}>CL_\pi$ and $E_{\rm EMC}$ together suppress about 75\% of background at the cost of 50\% of signal.
To suppress the backgrounds from the hadronic $D$ decays,
the maximum energy of any photon that is not used in the tag selection ($E_{\rm extra~\gamma}^{\rm max}$) is required to be less than 0.25~GeV,
and no additional charged track ($N^{\rm extra}_{\rm char}$) is allowed.

In the selection of $D^+\to\pi^+\pi^-\mu^+\nu_\mu$,
the energy deposited in the EMC of $\pi^+$ is required to be less than 0.8 times its reconstructed momentum in the MDC, to minimize the misidentified background from positrons.
To reject the background events associated with $K^0_S\to \pi^+_{\pi\to\mu}\pi^-$, where
$\pi^+_{\pi\to\mu}$ denotes a track of pion identified as a muon candidate,
the invariant mass of the $\mu^+\pi^-$ combination ($M_{\mu^+\pi^-}$) is required to be outside the interval $(0.446,\,0.518)$\,GeV$/c^{2}$.
To reduce the peaking backgrounds of $D^+\to\bar K^0\pi^+(\pi^0)$ and $D^+\to\pi^+\pi^+\pi^-(\pi^0)$, the invariant mass of the $\pi^+\pi^-\mu^+$ combination ($M_{\pi^+\pi^-\mu^+}$) is required to be less than 1.56~GeV/$c^2$.
To further suppress backgrounds associated with additional neutral pions, it is required that there is no extra $\pi^0$ besides that used in the SL selection ($N^{\rm extra}_{\pi^0}$),  which suppresses about of 36\% background at the cost of 4\% of signal.

\begin{figure*}[htbp]
\centering
\includegraphics[width=0.6\linewidth]{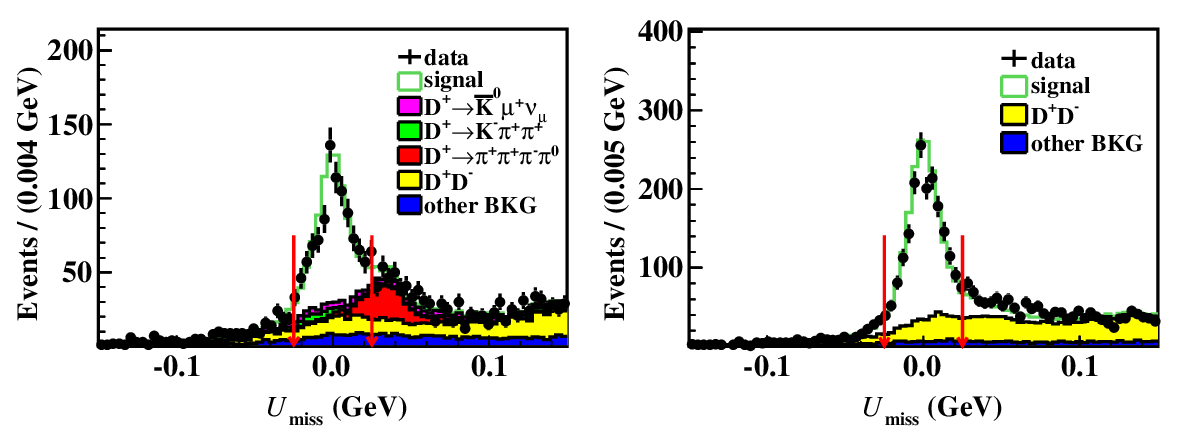}
\caption{\label{fig:fit1}
The $U_{\rm miss}$ distributions of the accepted candidates for
(left)  $D^+\to \pi^+\pi^-\mu^{+}\nu_{\mu}$ and (right) $D^+\to \pi^+\pi^-e^{+}\nu_e$.
The red arrows denote the signal region.
The events entering the distributions are required to lie outside $M_{\pi^+\pi^-}\in (0.461,0.533)$ GeV$/c^2$ for $D^+\to\pi^+\pi^-\mu^+\nu_{\mu}$, and $M_{\pi^+\pi^-}\in (0.42,0.56)$ GeV$/c^2$ for $D^+\to\pi^+\pi^-e^+\nu_{e}$.
}
\end{figure*}

To help  separate the SL signals from backgrounds, we define a kinematic quantity
of $U_{\rm miss}\equiv
E_{\mathrm{miss}}-|\vec{p}_{\mathrm{miss}}|$. Here, $E_{\mathrm{miss}}\equiv
E_{\mathrm{beam}}-E_{\pi^+\pi^-}-E_{\ell^{+}}$ and
$\vec{p}_{\mathrm{miss}}\equiv
\vec{p}_{D^+}-\vec{p}_{\pi^+\pi^-}-\vec{p}_{\ell^{+}}$ are the missing energy
and momentum of the SL event in the $e^+e^-$ center-of-mass frame, respectively, in
which $E_{\pi^+\pi^-(\ell^+)}$ and $\vec{p}_{\pi^+\pi^-(\ell^+)}$ are the energy
and momentum of the $\pi^+\pi^-(\ell^+$) candidates. The
$U_{\mathrm{miss}}$ resolution is improved using $\vec{p}_{D^+} \equiv
{-\hat{\vec{p}}_{D^-}}\cdot\sqrt{E_{\mathrm{beam}}^{2}-m_{D^+}^{2}}$, where
$\hat{\vec{p}}_{D^-}$ is the unit vector in the momentum direction of the tag
$D^-$ and $m_{D^+}$ is the nominal $D^+$ mass~\cite{pdg2022}.
The $U_{\rm miss}$ distribution of the correctly reconstructed signals
is expected to peak around zero, as shown in Fig.~\ref{fig:fit1}.
The events within $U_{\rm miss}\in(-0.025,0.025)$~GeV are retained for further analysis.
This requirement retains 83\% of the signal and removes 74\% of the all backgrounds for $D^+\to\pi^+\pi^-\mu^+\nu_\mu$, in which the major peaking background is from $D^+\to\pi^+\pi^-\pi^+\pi^0$; while retains 75\% of the signal and removes 82\% of the all backgrounds for $D^+\to\pi^+\pi^-e^+\nu_e$.

\section{Branching fractions}
\subsection{Signal efficiencies and branching fractions}

\begin{figure*}[htbp]
\centering
\includegraphics[width=0.6\linewidth]{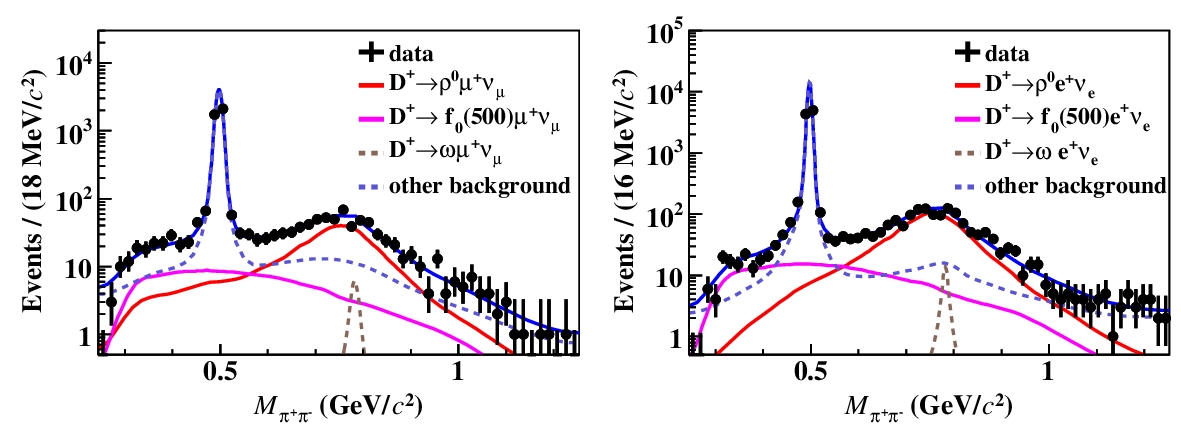}
\caption{\label{fig:fit2}
The $M_{\pi^+\pi^-}$ distributions of the candidates for
(left)  $D^+\to \pi^+\pi^-\mu^{+}\nu_{\mu}$ and (right) $D^+\to \pi^+\pi^-e^{+}\nu_e$ with the fit results overlaid (blue solid line).
The contributions of the signal and various background components are indicated, where `other BKG' includes all other components.}
\end{figure*}

Figure~\ref{fig:fit2} shows the $M_{\pi^+\pi^-}$ distributions of the accepted SL events in data.
According to the previous study of $D^+\to \pi^+\pi^-e^+\nu_e$~\cite{bes3-rhoev}, we ignore the non-resonant component except for
$f_0(500)$, $\rho^0$, and $\omega$. In this case, the signal yields from different components
are determined by an unbinned maximum likelihood fit to these distributions.
In the fit, the shapes of $D^+\to f_0(500)\ell^+\nu_\ell$, $D^+\to \rho^0\ell^+\nu_\ell$,
$D^+\to \omega \ell^+\nu_\ell$ are derived from individual signal MC samples
where the  Bugg function~\cite{ref29}, Gounaris-Sakurai~(GS) function~\cite{ref28} and GS$\times$RBW function~\cite{ref30} are used to describe the $f_0(500)$, $\rho^0$ and $\omega$ resonances, respectively.
Here RBW is a relativistic Breit-Wigner function with a constant width~\cite{ref30}.
The background shapes are derived from the inclusive MC sample.
MC studies show that the peaking backgrounds around 0.498 GeV/$c^2$ are mainly from the decays $D^+\to K^0_S\ell^+\nu_\ell$
and some hadronic $D$ decays containing $K^0_S$. In the fit, the peaking background component containing $K^0_S$ has been forced to have the same normalization as the
smoother background derived from the inclusive MC sample.
The amplitude analysis of $D\to \pi^+\pi^- e^+\nu_e$ in Ref. ~\cite{bes3-rhoev} shows that the interference between $D^+\to f_0(500)e^+\nu_e$ and $D^+\to \rho^0e^+\nu_e$ is negligible. With the same data sample, we take this conclusion into account to simplify the analysis.
The yields of $D^+\to \omega \ell^+\nu_\ell$ are fixed based on the BF previously measured by BESIII~\cite{bes3-rhoev}, while the yields of other components are free parameters.
The fit results on the $M_{\pi^+\pi^-}$ distributions are also shown in Fig.~\ref{fig:fit2}.
From these fits, we obtain the yields of $D^+\to \rho^0\ell^+\nu_\ell$ and $D^+\to f_0(500)\ell^+\nu_\ell$ decays.

\begin{figure*}[htbp]
	\begin{center}
		\subfigure{\includegraphics[width=0.9\linewidth]{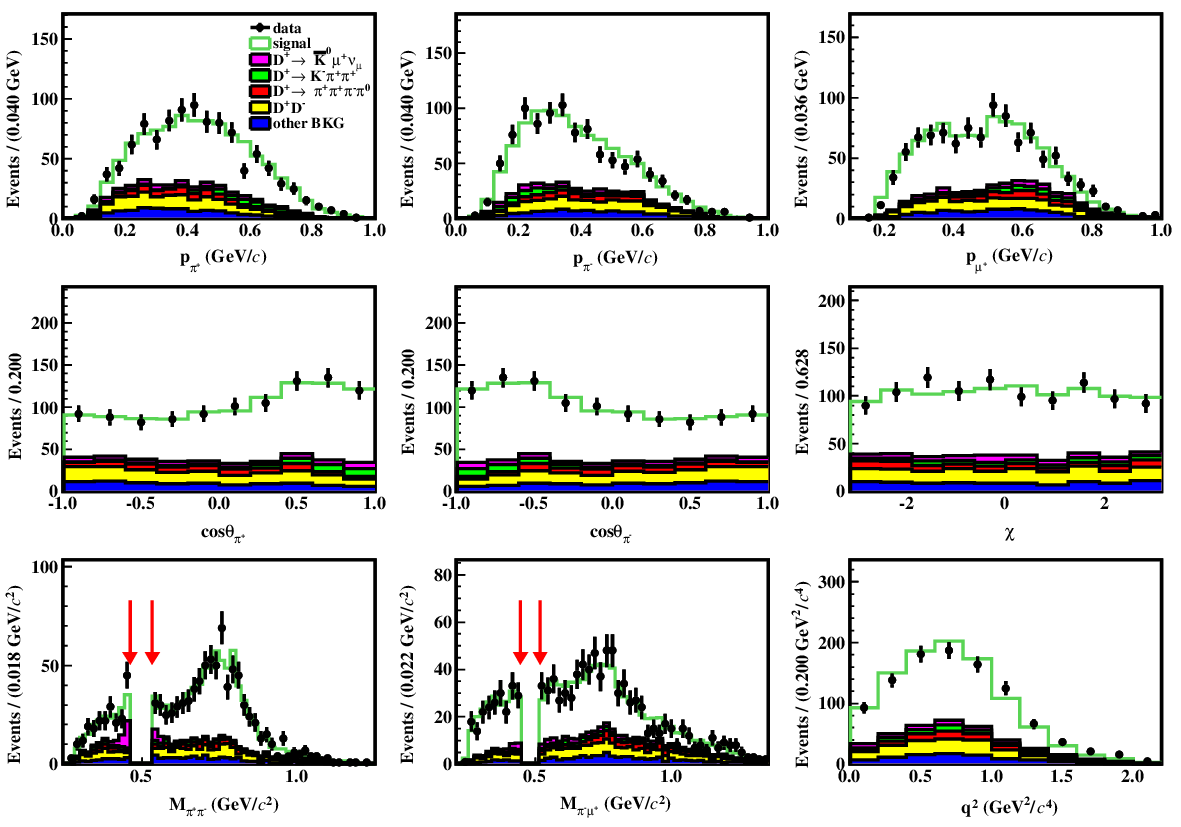}}
		\caption{
			The distributions of (top row) momentum and (middle row) $\cos\theta$ of $\pi^+$, $\pi^-$ and $\mu^+$, (bottom row) invariant-mass distributions of $\pi^+\pi^-$, $\pi^-\mu^+$ and $q^2$ distribution from the accepted candidates of $D^+\to \pi^+\pi^- \mu^+\nu_\mu$ for data and inclusive MC sample.
			$q^2$ is the mass square of the $\mu^+\nu_{\mu}$ system; $\cos\theta_{\pi^+}$ ($\cos\theta_{\pi^-}$), the helicity angle is the angle between the $\pi^+$ ($\pi^-$) and the mother decay plane in the $\pi^+\pi^-$ rest frame;
			$\chi$ is the coplanarity angle between the $\pi^+\pi^-$ and $\mu^+\nu_\mu$ decay planes.
			To enhance the signal distributions,
			an additional requirement of $M_{\pi^+\pi^-}\in (0.461,0.533)$ GeV$/c^2$ has been imposed in addition to the baseline selection criteria.
			\label{fig:mcdata}}
	\end{center}
\end{figure*}

The BF of the SL decay is determined by
\begin{equation}
\label{eq:bf}
{\mathcal B}_{\rm sig}=N_{\mathrm{obs}}/(N_{\mathrm{tag}}^{\rm tot}\cdot \varepsilon_{\rm sig}),
\end{equation}
where $N_{\rm tag}^{\rm tot}$ and $N_{\rm obs}$ are the tag and SL signal
yields in the data sample, and $\varepsilon_{\rm sig}=\Sigma_i [(\varepsilon^i_{\rm tag,sig}\cdot N^i_{\rm tag})/(\varepsilon^i_{\rm tag}\cdot N^{\rm tot}_{\rm tag})]$ is the efficiency of detecting the SL decay in the presence of the tag $D^-$ meson.  Here, $i$ denotes the tag mode, and $\varepsilon_{\rm tag}$ and $\varepsilon_{\rm tag,sig}$ are the efficiencies of selecting the tag and simultaneously selecting the tag and signal candidates, respectively.
Table~\ref{tab:bfs} summarizes the SL yields in data ($N_{\rm obs}$),  the statistical significances ($\mathcal S$), the signal efficiencies ($\varepsilon_{\rm sig}$),
and the obtained BFs (${\mathcal B}_{\rm sig}$).
For each signal decay, the statistical significance is calculated according to
$\sqrt{-2{\rm ln ({\mathcal L_0}/{\mathcal L_{\rm max}}})}$,
where ${\mathcal L}_{\rm max}$ and ${\mathcal L}_0$ are the maximum likelihoods of the fits with and without including the signal component, respectively.
The reliability of the efficiency determinations has been verified by
the comparison of the distributions of momenta and $\cos\theta$ of the $\pi^+$, $\pi^-$, and $\ell^+$ of
the selected $D^+\to\pi^+\pi^-\ell^+\nu_\ell$ candidate events between data
and MC simulation. As shown in Fig.~\ref{fig:mcdata}, a good consistency between data and MC simulation for $D^+\to \pi^+\pi^-\mu^+\nu_\mu$ is observed.
Similar comparison has also shown good data-MC consistence of these variables for $D^+\to \pi^+\pi^-e^+\nu_e$ in Ref.~\cite{bes3-rhoev}.

\begin{table}[htbp]\centering
  \caption{The signal yields in data, the statistical significances,
    the signal efficiencies and the absolute BFs
    of different signal modes. The BFs of $D^+\to f_0(500)\ell^+\nu_\ell$ include the BF of $f_0(500)\to\pi^+\pi^-$.
     The statistical uncertainties of the $\mu$ channels are worse than those of the $e$ channels by a factor of two mainly due to stricter selection criteria and much higher backgrounds. }
\label{tab:bfs}
\small
\begin{tabular}{lcccc}
\hline\hline
  Signal mode & $N_{\rm obs}$ & $\mathcal S$ ($\sigma$) & $\epsilon_{\rm sig}$ (\%) & ${\mathcal B}_{\rm sig} (\times10^{-3})$\\
  \hline
$f_0(500)\mu^{+}\nu_{\mu}$   & $ 209\pm38$ &$5.9$  & $18.93\pm0.13$ &$0.72\pm0.13$ \\
$\rho^0\mu^{+}\nu_{\mu}$     & $ 496\pm38$ &$>10$ & $19.86\pm0.13$ &$1.64\pm0.13$ \\
$f_0(500) e^{+}\nu_e$        & $ 412\pm43$ &$>10$  & $44.76\pm0.25$ &$0.60\pm0.06$ \\
$\rho^0 e^{+}\nu_e$          & $1237\pm47$ &$>10$ & $44.12\pm0.25$ &$1.84\pm0.07$ \\
\hline\hline
\end{tabular}
\end{table}

\subsection{Systematic uncertainties in branching fractions}

The systematic uncertainties in the BF measurements are discussed below.
The uncertainty in the total yield of tag $D^-$ mesons was previously assigned as 0.5\% in Refs.~\cite{epjc76,cpc40,bes3-pimuv}.
The systematic uncertainty in the tracking (PID) efficiency per $\pi^\pm$ is assigned as 0.2\% (0.3\%),
by analyzing the $D\bar D$ hadronic events~\cite{D-PP}.
The systematic uncertainties due to the tracking (PID) efficiencies per $\mu^+$ or $e^+$ are assigned as 0.2\% (0.5\%)
by using the $e^+e^-\to\gamma\mu^+\mu^-$ and $e^+e^-\to\gamma e^+e^-$ control samples,
respectively, where the data-MC differences of the two-dimensional (momentum
and $\cos\theta$) distributions of the control samples are
re-weighted to match those of the SL signal decays.
The uncertainty associated with the combined $E_{\rm extra~\gamma}^{\rm max}$, $N^{\rm extra}_{char}$ (and $N^{\rm extra}_{\pi^0}$) requirements is
taken to be 1.3\% by analyzing hadronic $D^+D^-$ candidate events.
The systematic uncertainty due to the $U_{\rm miss}$ requirement is estimated to be 0.5\% by
analyzing the control sample of $D^+\to K^0_S \ell^+\nu_\ell$.

To estimate the uncertainties due to the $M_{\mu^+\pi^-}$ requirements,
we remeasure the  BFs while varying the
$M_{\mu^+\pi^-}$ requirement in the range from (0.456,0.508) to (0.436,0.528) GeV/$c^2$,
with a step of 4 MeV/$c^2$ corresponding to  the approximate $K^0_S$ mass resolution.
To estimate the uncertainty due to the $M_{\pi^+\pi^-\mu^+}$ requirement,
we remeasure the BFs while varying the $M_{\pi^+\pi^-\mu^+}$ condition from 1.4 to 1.6 GeV/$c^2$, with a step size of 20 MeV/$c^2$.  Fitting the measurements obtained from this procedure with a linear function yields an average BF from this exercise.
The differences between the average BF and the baseline value are assigned as the systematic uncertainties,
which are 0.2\% and 0.5\% for the $M_{\mu^+\pi^-}$ and $M_{\pi^+\pi^-\mu^+}$ requirements, respectively.

The uncertainty associated with the $M_{\pi^+\pi^-}$ fit is estimated by examining the BF changes when using
a RBW signal description with Gaussian smearing of the MC-simulated signal shape, varying the amplitude analysis model parameters in their uncertainties 600 times randomly,
varying the assumed BFs of the peaking background components by $\pm 1\sigma$,
and varying the smooth parameters of other background contributions.
The uncertainties arising from the finite sample size of the signal MC  are 0.7\%, 0.6\% and 0.6\% for
$D^+\to f_0(500)\mu^+\nu_\mu$, $D^+\to\rho^0\mu^+\nu_\mu$, and
$D^+\to f_0(500)e^+\nu_e$, respectively.
The uncertainties due to the signal MC model for these three decays are 3.6\%, 3.6\% and 0.9\%, respectively,
which are the differences of the baseline signal efficiencies and those obtained with the signal MC samples after varying the values of the input FFs by $\pm 1\sigma$.

For each signal decay, adding these uncertainties in quadrature yields the total systematic uncertainty, which is 11.0\% for $D^+\to f_0(500)\mu^+\nu_\mu$, 6.4\% for $D^+\to\rho^0\mu^+\nu_\mu$, 7.7\% for $D^+\to f_0(500)e^+\nu_e$, and 3.5\% for $D^+\to\rho^0e^+\nu_e$.
Table~\ref{table:sys} summarizes the sources of the systematic uncertainties in the measurements of the BFs of $D^+\to f_0(500)\ell^+\nu_{\ell}$ and $D^+\to \rho^0\ell^+\nu_{\ell}$.

\begin{table*}[htbp]
	\centering
	\caption{Relative systematic uncertainties (in \%) in the measurements of the BFs of $D^+\to f_0(500)\ell^+\nu_{\ell}$ and $D^+\to \rho^0\ell^+\nu_{\ell}$. \label{table:sys}}
	\begin{tabular}{l|cc|cc}
		\hline
		\hline
		Source  & $D^+\to\rho^0\mu^+\nu_{\mu}$&$D^+\to f_0(500)\mu^+\nu_\mu$& $D^+\to\rho^0 e^+\nu_{e}$& $D^+\to f_0(500)e^+\nu_e$\\
		\hline
		$N_{\rm tag}^{\rm tot}$               &0.5     &0.5      &0.5     &0.5 \\
		$\pi^\pm$ tracking                  &0.4     &0.4      &0.4     &0.4\\
		$\pi^\pm$ PID                       &0.6     &0.6      &0.6     &0.6\\
		$\ell^+$ tracking                    &0.2     &0.2      &0.2     &0.2\\
		$\ell^+$ PID                         &0.5     &0.5      &0.5     &0.5\\
		$U_{\rm miss}$ requirement            &0.5        &0.5         &0.5        &0.5   \\
		$E_{\rm extra\gamma}^{\rm max}\& N^{\rm extra}_{\rm char}$      &1.3  &1.3  &1.3  &1.3\\
		$M_{\pi^-\mu^+}$ requirement           &0.2     &0.2      &$-$      &$-$\\
		$M_{\pi^+\pi^-\mu^+}$ requirement      &0.5     &0.5      &$-$      &$-$\\
		$M_{\pi^+\pi^-}$ fit                &4.9      &10.2     &2.8     &7.4    \\
		MC sample size                       &0.6      &0.7      &0.6     &0.6 \\
		MC model                            &3.6     &3.6       &0.9     &0.9\\
		\hline
		Total                               &6.4     &11.0      &3.5     &7.7\\
		\hline
		\hline
	\end{tabular}
\end{table*}

\subsection{Results and discussion}
Finally, the BFs of $D^+\to f_0(500)\mu^+\nu_\mu$ with $f_0(500)\to \pi^+\pi^-$ and $D^+\to \rho^0\mu^+\nu_\mu$ are determined to be ${\mathcal B}_{D^+\to f_0(500)\mu^+\nu_\mu}\times{\mathcal B}_{f_0(500)\to\pi^+\pi^-}=(0.72\pm0.13_{\rm stat}\pm0.08_{\rm syst})\times 10^{-3}$ and ${\mathcal B}_{D^+\to \rho^0 \mu^+\nu_\mu}=(1.64\pm0.13_{\rm stat}\pm0.10_{\rm syst})\times 10^{-3}$.
	As cross checks, we have also determined the BFs of $D^+\to f_0(500) e^{+}\nu_e$, $f_0(500)\to\pi^+\pi^-$ and $D^+\to \rho^0 e^{+}\nu_e$ to be ${\mathcal B}_{D^+\to f_0(500) e^+\nu_e}\times{\mathcal B}_{f_0(500)\to\pi^+\pi^-}=(0.60\pm0.06_{\rm stat}\pm0.05_{\rm syst})\times 10^{-3}$ and ${\mathcal B}_{D^+\to \rho^0 e^+\nu_e}=(1.84\pm0.07_{\rm stat}\pm0.06_{\rm syst})\times 10^{-3}$. They are
    consistent with those from the amplitude analysis of $D^+\to \pi^+\pi^- e^{+}\nu_e$~\cite{bes3-rhoev}.
	Combining the ${\mathcal B}_{D^+\to h\mu^+\nu_\mu}$ ($h=f_0(500)$ or $\rho^0$) obtained in this work and the ${\mathcal B}_{D^+\to h e^+\nu_e}$ from a previous BESIII analysis~\cite{bes3-rhoev} gives the BF ratios
	${\mathcal B}_{D^+\to \rho^0 \mu^+\nu_\mu}/{\mathcal B}_{D^+\to \rho^0 e^+\nu_e}=0.88\pm0.10$ and
	${\mathcal B}_{D^+\to f_0(500) \mu^+\nu_\mu}/{\mathcal B}_{D^+\to f_0(500) e^+\nu_e}=1.14\pm0.26$,
	which are compatible with the predictions of 0.96~\cite{rho_CCQM,rho_LCSR} and 0.89$\sim$0.91~\cite{RMWang2023,f0_bf}, respectively, made under the assumption of lepton universality.

\section{Hadronic form factors}
\subsection{Hadronic form factors}

\begin{figure*}[htbp]\centering
	\includegraphics[width=0.9\linewidth]{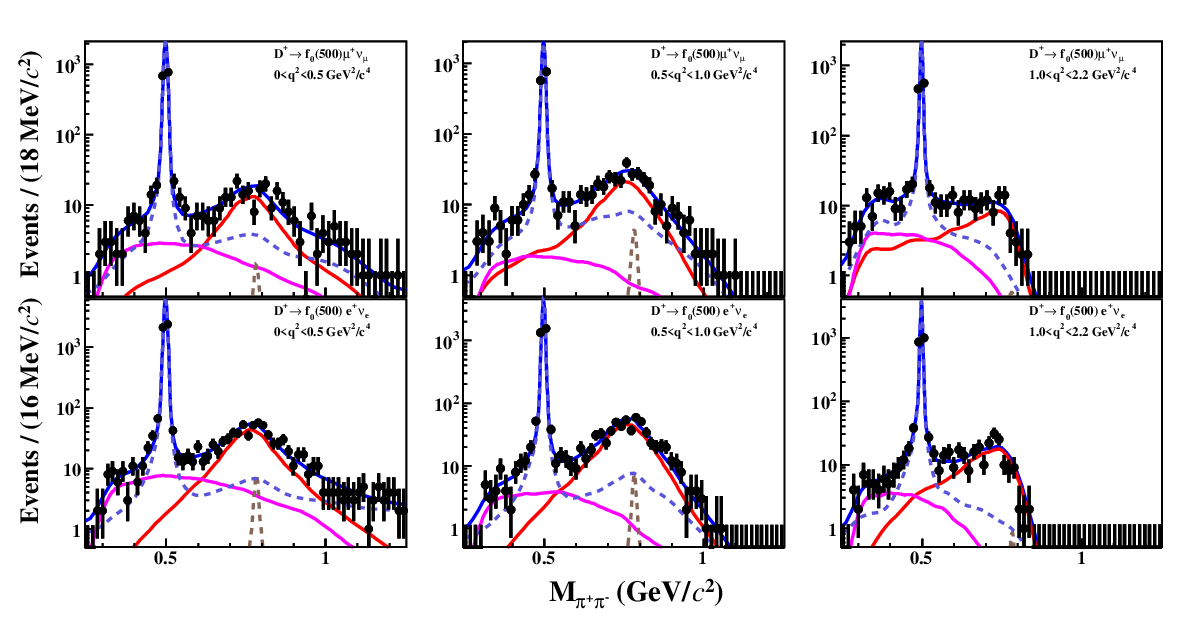}
	\caption{Fits to the $M_{\pi^+\pi^-}$ distributions of the $D^+\to f_0(500)\ell^+\nu_{\ell}$ candidate events. Data are shown as dots with
		error bars. The blue solid curves are the fit results; the red solid line and pink solid lines denote the SL signals of $D^+\to\rho^0\ell^+\nu_{\ell}$ and $D^+\to f_0(500)\ell^+\nu_{\ell}$, respectively; the black dotted curves are the peaking background of $D^+\to\omega\ell^+\nu_{\ell}$; and the blue dotted curves are the fitted combinatorial backgrounds.}
	\label{fig:fiteachbina}
\end{figure*}

We study the $D^+\to f_0(500) \ell^+\nu_\ell$ decay dynamics, by dividing the SL
candidate events into three $q^2$ intervals: $(0.0,0.5)$, $(0.5,1.0)$, and $(1.0,2.2)$~GeV$^2$/$c^4$, where $q^2$ is the four-momentum transfer square of the $\ell^+\nu_\ell$ system. A least-$\chi^2$ fit is performed to the measured ($\Delta\Gamma_{\rm msr}^i$) and theoretically expected ($\Delta\Gamma_{\rm th}^i$) partial decay rates in the $i$th $q^2$ interval. The $\Delta\Gamma^{i}_{\rm msr}$ is determined by
\begin{equation} \label{eq2}
	\Delta\Gamma^{i}_{\rm msr}=\sum_{j}^{3}(\varepsilon^{-1})_{ij}N_{\mathrm{obs}}^{j}/(\tau_{D^+} \cdot N_{\mathrm{tag}}^{\rm tot}),
\end{equation}
where $N_{\rm obs}^j$ is the SL signal yield in the $j$th $q^{2}$ interval in data, $\tau_{D^+}$ is the $D^+$ lifetime, and $\varepsilon$ is the efficiency matrix.
Figure~\ref{fig:fiteachbina} shows the results of the fits to the $M_{\pi^+\pi^-}$ distributions in the reconstructed $q^2$ bins, using the same fitting method as the data.
Table~\ref{table:eff} gives the weighted efficiency matrices of $D^+\to f_0(500)\ell^+\nu_{\ell}$, which have been averaged over all six tag modes.
Table~\ref{table:detf} presents  the numbers of reconstructed events in data $N^i_{\rm obs}$ obtained by fitting the $M_{\pi^+\pi^-}$ distributions, the numbers of produced events $N^i_{\rm pro}$, and the measured partial decay rates $\Delta\Gamma^i_{\rm msr}$ in various $q^2$ intervals for $D^+\to f_0(500)\ell^+\nu_{\ell}$.

\begin{table*}[htbp]
	\centering
	\caption{The efficiency matrices of $D^+\to f_0(500)\ell^+\nu_{\ell}$ averaged over six tag modes, where $\varepsilon_{ij}$ represents the efficiency of events produced in the $j$-th $q^2$ interval and reconstructed in the $i$-th $q^2$ interval.
		\label{table:eff}}
	\begin{tabular}{l|ccc|ccc}
		\hline
		\hline
		& &$D^+\to f_0(500)\mu^+\nu_{\mu}$& & &$D^+\to f_0(500)e^+\nu_{e}$ &\\
		$\varepsilon_{ij}$ (\%) &1&2&3&1&2&3\\
		\hline
		1	&15.18 &0.83  &0.41  &45.33 &0.96  &0.02	\\
		2	&1.31  &19.01 &0.98  &0.45  &42.45 &1.10	\\
		3   &1.05  &0.89  &17.17 &0.02  &0.55  &40.11  \\
		\hline
		\hline
	\end{tabular}
\end{table*}
\begin{table*}[htbp]
	\centering
	\caption{The partial decay rates of $D^+\to f_0(500)\ell^+\nu_{\ell}$ in various $q^2$ intervals.
		\label{table:detf}}
	\begin{tabular}{l|ccc|ccc}
		\hline
		\hline
		& &$D^+\to f_0(500)\mu^+\nu_{\mu}$& & &$D^+\to f_0(500)e^+\nu_{e}$ &\\
		$i$ &1&2&3&1&2&3\\
		$q^2~({\rm GeV}^2/c^4)$& $(0.0, 0.5)$& (0.5, 1.0)& (1.0, 2.2)&$(0.0, 0.5)$& (0.5, 1.0)& (1.0, 2.2)\\
		\hline
		$N^i_{\rm obs}$&$74\pm21 $&$45\pm22 $&$72\pm22 $&$221\pm31 $&$95\pm21 $&$63\pm16 $\\
		$N^i_{\rm pro}$&$469\pm136$&$187\pm115$&$380\pm131$&$483\pm68$&$215\pm49$&$154\pm39$\\
		$\Delta\Gamma^i_{\rm msr}({\rm ns}^{-1})$&$0.298\pm0.087$&$0.119\pm0.073$&$0.242\pm0.083$&$0.307\pm0.043  $&$0.137\pm0.031  $&$0.098\pm0.025$  \\
		\hline
		\hline
	\end{tabular}
\end{table*}

The $\Delta\Gamma^{i}_{\rm th}$ is expressed as~\cite{zsl51,ddr}
\begin{widetext}
\begin{equation}
\begin{array}{l}
	\displaystyle \Delta\Gamma^{i}_{\mathrm{th}} =
   \int_{q_{\mathrm{min}(i)}^{2}}^{q_{\mathrm{max}(i)}^{2}}\int_{s_{\mathrm{min}}}^{s_{\mathrm{max}}}\frac{G_{F}^{2}|V_{cd}|^{2}}{192\pi^{4}m^3_{D^+}}\left(1-\frac{m^{2}_{\ell}}{q^{2}}\right)^2\left(1+\frac{m^{2}_{\ell}}{2q^{2}}\right)\lambda^{\frac{3}{2}}(m^2_{D^+},s,q^2)|f^{f_0}_{+}(q^{2})|^{2}P(s){\rm d}s{\rm d}q^{2},\\
\end{array}
\end{equation}
\end{widetext}
where $G_{F}$ is the Fermi constant, $|V_{cd}|$ is the magnitude of the Cabibbo Kobayashi Maskawa matrix element governing $c \to d$ transition, $m_{D^+}$ is the mass of $D^+$, $m_{\ell}$ is the mass of lepton, $f_0$ denotes the  $f_0(500)$ meson, $s$ is the mass-squared of the $\pi^+\pi^-$ pair,
$\lambda(x,y,z)=x^2+y^2+z^2-2xy-2xz-2yz$,
and the $P(s)$ is formed with the Bugg resonant lineshape.

In this analysis, the hadronic FF ($f^{f_0}_{+}(q^{2})$) is parameterized by a two-parameter series expansion~\cite{Becher:2005bg} due to very limited statistics.
The series expansion is given as \cite{Becher:2005bg}
\begin{equation}
	\label{equation:eq4}
	f_{+}(t)=\frac{1}{P(t)\Phi(t,t_{0})}a_{0}(t_{0})(1+\sum_{k=1}^{\infty}r_{k}(t_{0})[z(t,t_{0})]^{k}),
\end{equation}
where
\begin{equation}
	\label{equation:eq5}
	\begin{array}{l}
		\displaystyle z(t,t_{0})=\frac{\sqrt{t_{+}-t}-\sqrt{t_{+}-t_{0}}}{\sqrt{t_{+}-t}+\sqrt{t_{+}-t_{0}}},\\
		\displaystyle t_{\pm}=(m_{D^+}\pm m_{f_0})^{2},\\
		\displaystyle t_{0}=t_{+}(1-\sqrt{1-t_{-}/t_{+}}).
	\end{array}
\end{equation}
 The $m_{f_0}$, mass of $f_0(500)$ meson, is taken as $(538\pm12)~{\rm MeV}/c^2$~\cite{MWf0}. The function $P(t)=z(t,m_{D^{*}}^{2})$ and $\Phi$ is given by
\begin{equation}
	\label{equation:eq6}
	\begin{array}{l}
		\displaystyle \Phi(t,t_{0})=\sqrt{\frac{1}{24\pi\chi_{V}}}(\frac{t_{+}-t}{t_{+}-t_{0}})^{1/4}(\sqrt{t_{+}-t}+\sqrt{t_{+}})^{-5}\\
		\displaystyle \times(\sqrt{t_{+}-t}+\sqrt{t_{+}-t_{0}})(\sqrt{t_{+}-t}+\sqrt{t_{+}-t_{-}})^{3/2}\\
		\displaystyle \times(t_{+}-t)^{3/4},
	\end{array}
\end{equation}
where $\chi_{V}$ can be obtained from dispersion relations using perturbative QCD and depends on the ratio of the $d$-quark mass over $c$-quark mass, $u=m_{d}/m_{c}$~\cite{ff4}. At the leading order, with $u=0$,
\begin{equation}
	\label{equation:eq7}
	\chi_{V}=\frac{3}{32\pi^{2}m_{c}^{2}}.
\end{equation}
Due to limited sample size in this analysis, we select the $k = 1$ case, which gives
\begin{equation}
	\label{equation:eq8}
	f_{+}(t)=\frac{1}{P(t)\Phi(t,t_{0})}\frac{f_{+}(0)P(0)\Phi(0,t_{0})}{1+r_1(t_0)z(0,t_0)}(1+r_1(t_0)z(t,t_0)).	
\end{equation}

\begin{figure*}[htbp] \centering
\includegraphics[width=0.6\linewidth]{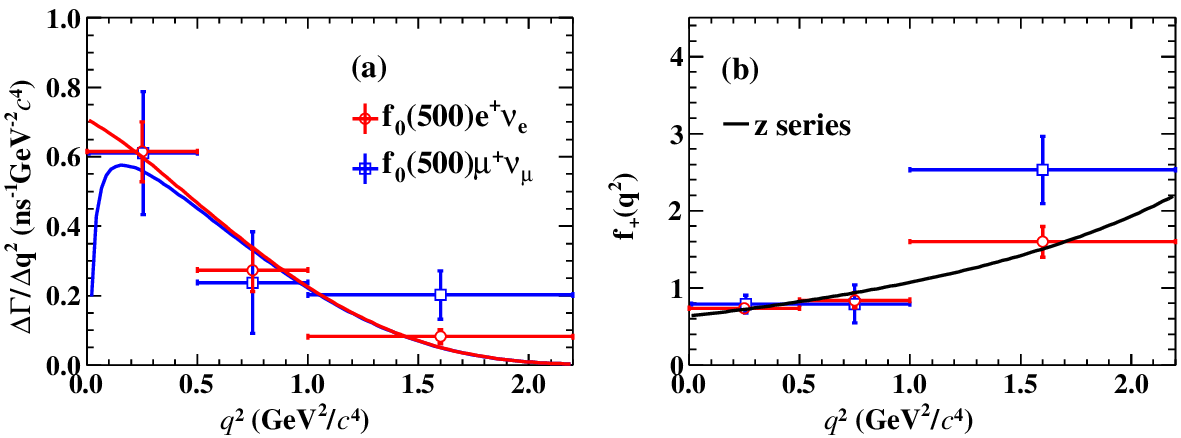}
\caption{ (a) The partial decay rates of $D^+\to f_0(500)\ell^+\nu_\ell$ and (b) projections to
  $f_+^{f_0}(q^2)$ with fit results overlaid, where the dots with error bars are data and the curves are the fit results.
  }
\label{fig:ddr}
\end{figure*}

A simultaneous fit is performed on the differential decay rates of $D^+\to f_0(500)\mu^+\nu_\mu$ and $D^+\to f_0(500)e^+\nu_e$,
in which the two decays are constrained to have the same parameters for the hadronic FF.
The fit results and the projections on $f^{f_0}_+(q^2)$ are shown in Fig.~\ref{fig:ddr}.
The baseline fit parameters are taken from the fit performed with the combined statistical and systematic covariance matrix ($\rho_{ij}^{\rm stat}$ and $\rho_{ij}^{\rm syst}$),
and their statistical uncertainties are taken from the fit only with the statistical covariance matrix.
Tables~\ref{table:stat} and \ref{table:sysm} summarize the statistical and systematic covariance matrices ($\rho^{\rm stat}_{ij}$ and $\rho^{\rm syst}_{ij}$) for the measured partial decay rates of $D^+\to f_0(500)\ell^+\nu_\ell$ in different $q^2$ intervals, respectively. The upper-left $3\times3$ matrix represents the result of $D^+\to f_0(500)\mu^+\nu_\mu$ channel, the lower-right $3\times3$ matrix represents the result of $D^+\to f_0(500)e^+\nu_e$ channel, and the upper-right $3\times3$ matrix represents the correlation between the two channels in different $q^2$ intervals.

\begin{table*}[htbp]
	\centering
	\caption{Statistical matrices for the measured partial decay rates of $D^+\to f_0(500)\ell^+\nu_{\ell}$ in different $q^2$ intervals.
		\label{table:stat}}
	\begin{tabular}{l|ccc|ccc}
		\hline
		\hline
		& &$D^+\to f_0(500)\mu^+\nu_{\mu}$& & &$D^+\to f_0(500)e^+\nu_{e}$ &\\
		$\rho^{\rm stat}_{ij}$&1&2&3&1&2&3\\
		\hline
		1	&1.000 &-0.121 &-0.081  	&0.000 &0.000 &0.000\\
		2	&       &1.000  &-0.094	    &0.000 &0.000 &0.000\\
		3   &       &       &1.000      &0.000 &0.000 &0.000\\
		1   &       &       &           &1.000 &-0.030 &-0.000\\
		2   &       &       &           &      &1.000  &-0.038	\\
		3   &       &       &           &      &       &1.000\\
		\hline
		\hline
	\end{tabular}
\end{table*}

\begin{table*}[htbp]
	\centering
	\caption{Systematic density matrices of the measured partial decay rates of $D^+\to f_0(500)\ell^+\nu_{\ell}$ in different $q^2$ intervals.
		\label{table:sysm}}
	\begin{tabular}{l|ccc|ccc}
		\hline
		\hline
		& &$D^+\to f_0(500)\mu^+\nu_{\mu}$& & &$D^+\to f_0(500)e^+\nu_{e}$ &\\
		$\rho^{\rm syst}_{ij}$ &1&2&3&1&2&3\\
		\hline
1       &1.000  &0.985  &0.679  &0.002  &0.002  &0.003  \\
2       &  &1.000  &0.611  &0.002  &0.002  &0.002  \\
3       &  &  &1.000  &0.004  &0.004  &0.005  \\
1       &  &  &  &1.000  &0.994  &0.879  \\
2       &  &  &  &  &1.000  &0.880  \\
3       &  &  &  &  &  &1.000  \\
		\hline
		\hline
	\end{tabular}
\end{table*}
For each parameter, the systematic uncertainty is obtained by calculating the quadratic difference of the uncertainties between these two fits.
The systematic uncertainties related to $\Delta\Gamma_{\rm msr}^{i}$ is estimated to be 4.4\% by the method in Ref.~\cite{bes3-D02kenu}, which is the same as the measurement content of the systematic uncertainties of BFs as in Table~\ref{table:sys}.
The uncertainty associated with $\Delta\Gamma^{i}_{\rm th}$ is assigned as 3.0\% by changing the parameters of Bugg resonant lineshape~\cite{ref29}.
The total systematic uncertainty is taken to be the square root of the quadratic sum of these contributions.

\subsection{Results and discussion}

\begin{figure*}[htbp]
\centering
\includegraphics[width=0.6\linewidth]{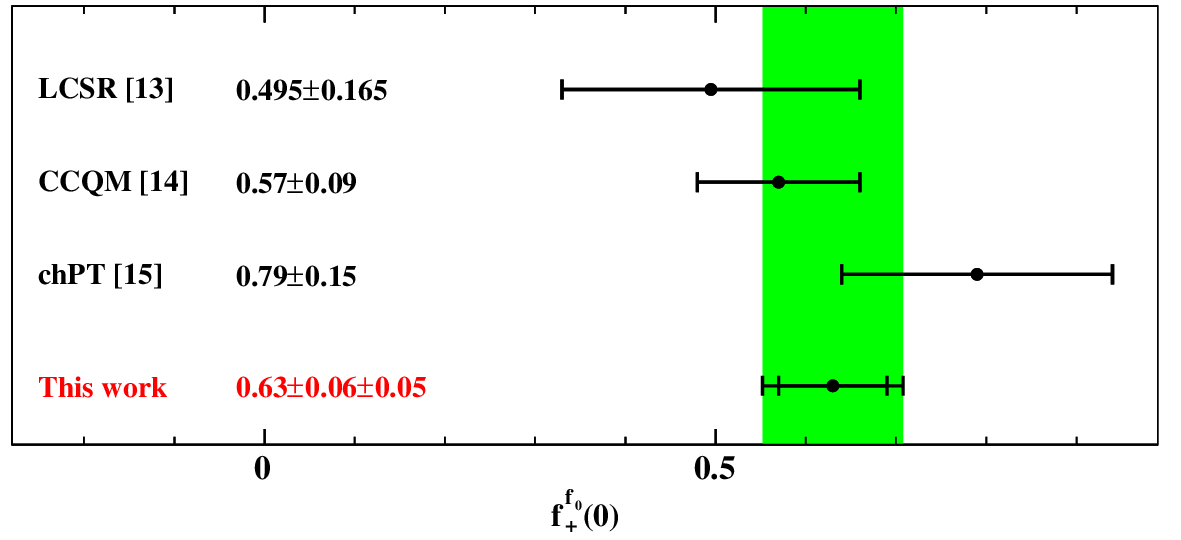}
\caption{\label{fig:com_ff}
The comparison of $f^{f_0}_{+}(0)$ obtained in this work with theoretical calculations. The form factors are shown as dots with
error bars.
}
\end{figure*}

From the fit, we determine $f_{+}^{f_0} (0)|V_{cd}|=0.143\pm0.014_{\rm stat}\pm0.011_{\rm syst}$ for the first time,
by analyzing simultaneously the partial decay rates of $D^+\to f_0(500) \mu^+\nu_\mu$ and $D^+\to f_0(500) e^+\nu_e$.
Using the value of $|V_{cd}|=0.22438\pm0.00044$ given in Ref.~\cite{pdg2022} yields
$f_{+}^{f_0}(0)=0.63\pm0.06_{\rm stat}\pm0.05_{\rm syst}$. We obtain the parameter of hadronic FF $r_1=-5.8\pm2.4\pm0.5$ and the correlation coefficient of the two parameters is 0.83. The fit quality is $\chi^2/ndf=5.3/4$, where $ndf$ is the number of degrees of freedom. The obtained hadronic FF is consistent with the prediction given by QCD light-cone sum rules (LCSR)~\cite{Dosch:2002rh}, covariant confined quark model (CCQM)~\cite{Gatto2000} and chiral perturbation theory (chPT)~\cite{chPT}, as shown in Fig.~\ref{fig:com_ff}. The obtained results are helpful for deepening and understanding of the nature of the $f_0(500)$ meson.

\section{Summary}

In summary, by analyzing 2.93~fb$^{-1}$ of $e^+e^-$ collision data taken at $\sqrt s=$ 3.773~GeV with
the BESIII detector,
we report the first observation of $D^+\to f_0(500)\mu^+ \nu_{\mu}$ with statistical significance of $5.9\sigma$. The absolute BFs of $D^+\to \rho^0\ell^+\nu_\ell$ and $D^+\to f_0(500)\ell^+\nu_\ell$ are measured.
As cross checks, we have also presented the BFs of $D^+\to f_0(500) e^{+}\nu_e$ with $f_0(500)\to\pi^+\pi^-$
and $D^+\to \rho^0 e^{+}\nu_e$ with $\rho^0\to\pi^+\pi^-$ , which are in agreement with the previous BESIII results~\cite{bes3-rhoev}.
From simultaneous fit to the partial decay rates of $D^+\to f_0(500) \mu^+\nu_\mu$ and $D^+\to f_0(500) e^+\nu_e$, the product of $f_{+}^{f_0}(0)|V_{cd}|$ is determined. Furthermore, taking the value of $|V_{cd}|$ from a standard model fit~\cite{pdg2022} as input, the form factor at zero momentum transfer squared $f_{+}^{f_0}(0)$ is determined. The measured hadronic FF is important to test the various theoretical calculations~\cite{Dosch:2002rh,Gatto2000,chPT}

\section{acknowledgments}

The BESIII Collaboration thanks the staff of BEPCII and the IHEP computing center for their strong support. This work is supported in part by National Key R\&D Program of China under Contracts Nos. 2023YFA1606000, 2020YFA0406400, 2020YFA0406300; National Natural Science Foundation of China (NSFC) under Contracts Nos. U1932102, 11635010, 11735014, 11835012, 11935015, 11935016, 11935018, 11961141012, 12025502, 12035009, 12035013, 12061131003, 12192260, 12192261, 12192262, 12192263, 12192264, 12192265, 12221005, 12225509, 12235017; Natural Science Foundation of Hunan Province, China under Contract No.~2021JJ40036 and the Fundamental Research Funds for the Central Universities under Contract No. 020400/531118010467; the Chinese Academy of Sciences (CAS) Large-Scale Scientific Facility Program; the CAS Center for Excellence in Particle Physics (CCEPP); Joint Large-Scale Scientific Facility Funds of the NSFC and CAS under Contract No. U1832207; CAS Key Research Program of Frontier Sciences under Contracts Nos. QYZDJ-SSW-SLH003, QYZDJ-SSW-SLH040; 100 Talents Program of CAS; The Institute of Nuclear and Particle Physics (INPAC) and Shanghai Key Laboratory for Particle Physics and Cosmology; European Union's Horizon 2020 research and innovation programme under Marie Sklodowska-Curie grant agreement under Contract No. 894790; German Research Foundation DFG under Contracts Nos. 455635585, Collaborative Research Center CRC 1044, FOR5327, GRK 2149; Istituto Nazionale di Fisica Nucleare, Italy; Ministry of Development of Turkey under Contract No. DPT2006K-120470; National Research Foundation of Korea under Contract No. NRF-2022R1A2C1092335; National Science and Technology fund of Mongolia; National Science Research and Innovation Fund (NSRF) via the Program Management Unit for Human Resources \& Institutional Development, Research and Innovation of Thailand under Contract No. B16F640076; Polish National Science Centre under Contract No. 2019/35/O/ST2/02907; The Swedish Research Council; U. S. Department of Energy under Contract No. DE-FG02-05ER41374.

\clearpage
\section{appendix}
\begin{figure*}[htbp]
	\centering
	\includegraphics[width=0.6\linewidth]{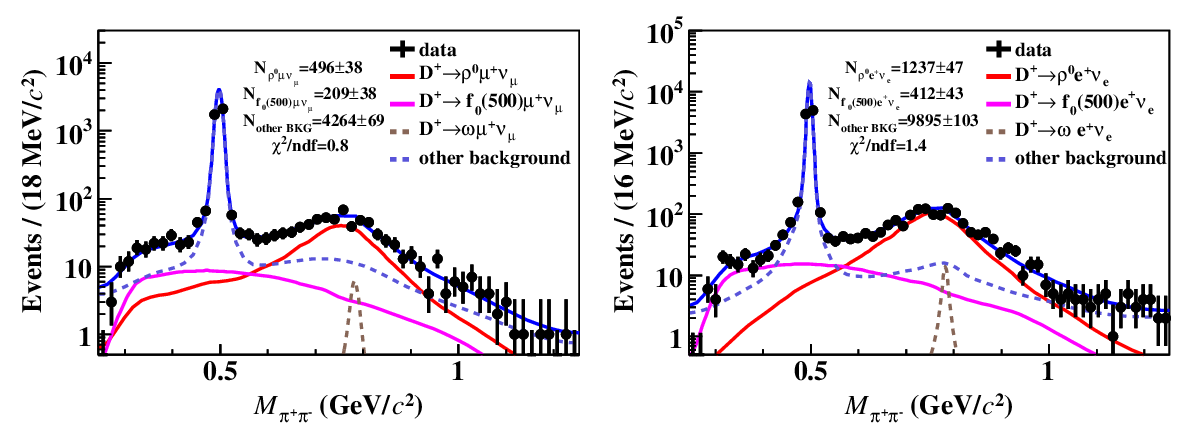}
	\caption{\label{fig:fit3}
		The $M_{\pi^+\pi^-}$ distributions of the candidates for
		(left)  $D^+\to \pi^+\pi^-\mu^{+}\nu_{\mu}$ and (right) $D^+\to \pi^+\pi^-e^{+}\nu_e$ with the nominal fit results overlaid (blue solid line).
		The contributions of the signal and various background components are indicated, where other background includes all other components.}
\end{figure*}

Figure~\ref{fig:fit3} shows the results of the fits to the $M_{\pi^+\pi^-}$ distributions, in which the signal shapes are derived from individual signal MC samples, and the background shapes are derived from the inclusive MC sample.

\begin{figure*}[htbp]
	\centering
	\includegraphics[width=0.6\linewidth]{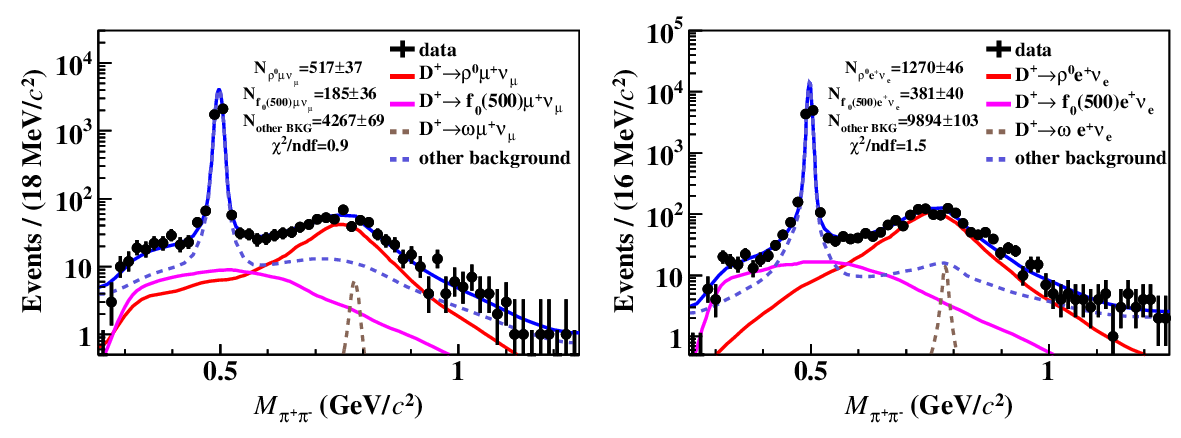}
	\caption{\label{fig:fit4}
		Fits to the $M_{\pi^+\pi^-}$ distributions of the candidates for
		(left)  $D^+\to \pi^+\pi^-\mu^{+}\nu_{\mu}$ and (right) $D^+\to \pi^+\pi^-e^{+}\nu_e$ with RBW as alternative signal shapes (blue solid line).
		The contributions of the signal and various background components are indicated, where other background includes all other components.}
\end{figure*}

Figure~\ref{fig:fit4} shows the results of the fits to the $M_{\pi^+\pi^-}$ distributions,
in which the shapes the $D^+\to f_0(500)\mu^+\nu_\mu$ and $D^+\to f_0(500)e^+\nu_e$ signals are described with the alternative
RBW shapes.

\begin{figure*}[htbp]
	\centering
	\includegraphics[width=0.6\linewidth]{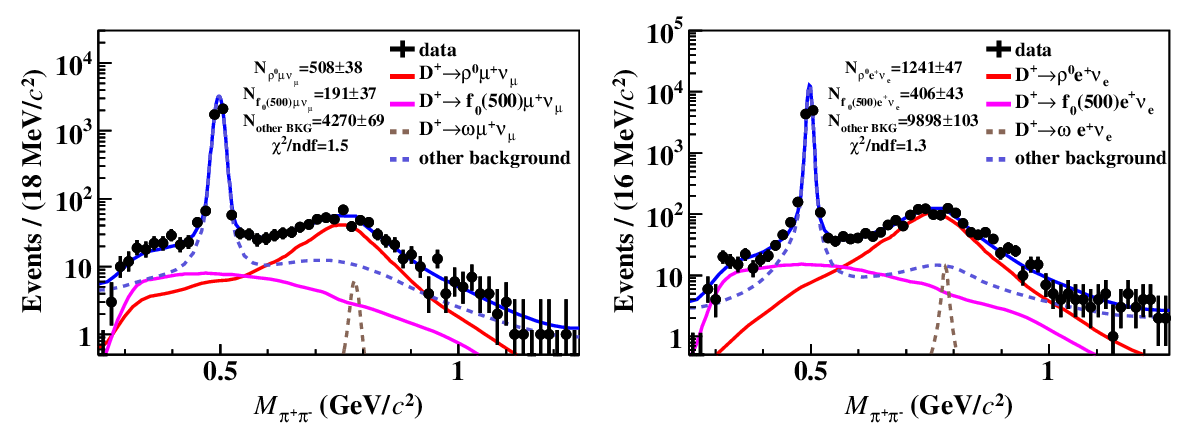}
	\caption{
		Fits to the $M_{\pi^+\pi^-}$ distributions of the candidates for
		(left) $D^+\to \pi^+\pi^-\mu^{+}\nu_{\mu}$ and (right) $D^+\to \pi^+\pi^-e^{+}\nu_e$,
   in which the parameter of other background shape is increased  by 0.5.
   	The contributions of the signal and various background components are indicated, where other background includes all other components.
	}
	\label{fig:hua2}
\end{figure*}

\begin{figure*}[htbp]
	\centering
	\includegraphics[width=0.6\linewidth]{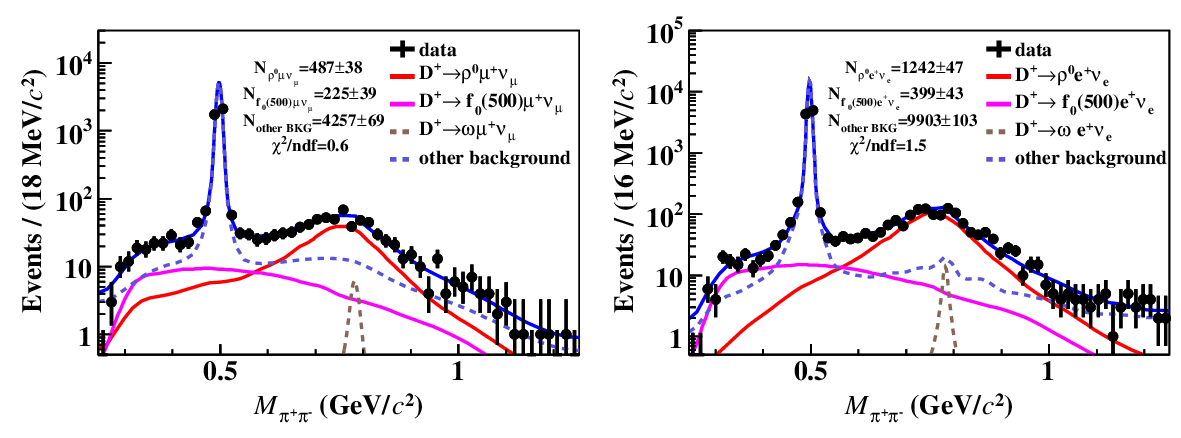}
	\caption{
		Fits to the $M_{\pi^+\pi^-}$ distributions of the candidates for
		(left) $D^+\to \pi^+\pi^-\mu^{+}\nu_{\mu}$ and (right) $D^+\to \pi^+\pi^-e^{+}\nu_e$,
   in which the parameter of other background shape is decreased  by 0.5.
		The contributions of the signal and various background components are indicated, where other background includes all other components.
	}
	\label{fig:hua2_e}
\end{figure*}

Figures~\ref{fig:hua2} and \ref{fig:hua2_e} show the fits to the $M_{\pi^+\pi^-}$ distributions of $D^+\to f_0(500)\mu^+\nu_\mu$,
in which the parameters of other background shapes are increased and decreased by 0.5, respectively.

\begin{figure*}[htbp]
	\centering
	\includegraphics[width=0.9\linewidth]{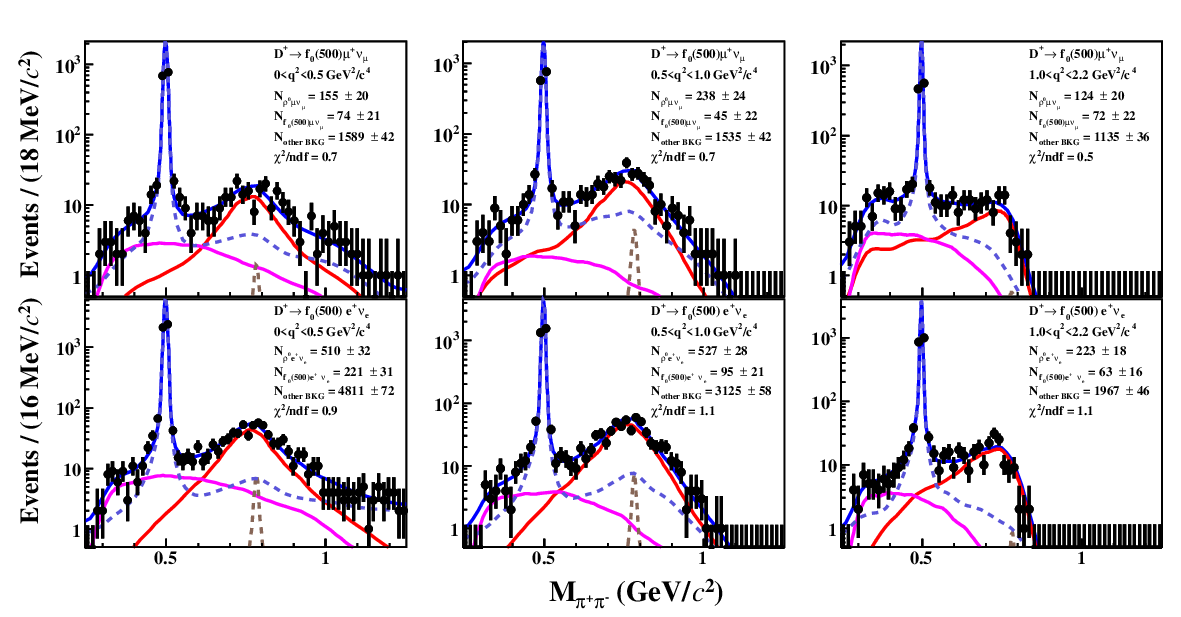}
	\caption{\label{fig:fit5}
		The $M_{\pi^+\pi^-}$ distributions of the candidates for
		(up)  $D^+\to \pi^+\pi^-\mu^{+}\nu_{\mu}$ and (down) $D^+\to \pi^+\pi^-e^{+}\nu_e$ in different reconstructed $q^2$ bins with the nominal fit results overlaid (blue solid line).
		The contributions of the signal and various background components are indicated, where other background includes all other components.}
\end{figure*}

Figure~\ref{fig:fit5} shows the results of the fits to the $M_{\pi^+\pi^-}$ distributions in different reconstructed $q^2$ bins, in which all the signal shapes are derived from individual signal MC samples, and the background shapes are derived from the inclusive MC sample.

\begin{figure*}[htbp]
	\centering
	\includegraphics[width=0.9\linewidth]{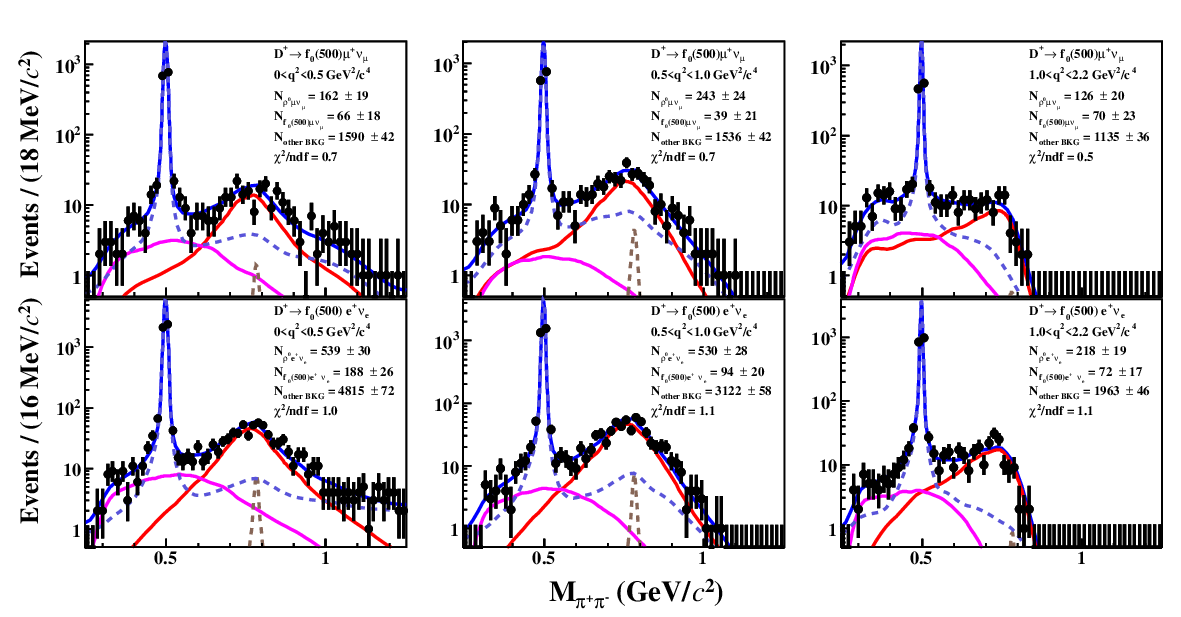}
	\caption{\label{fig:fit6}
		Fits to the $M_{\pi^+\pi^-}$ distributions of the candidates for
		(up)  $D^+\to \pi^+\pi^-\mu^{+}\nu_{\mu}$ and (down) $D^+\to \pi^+\pi^-e^{+}\nu_e$ with RBW as alternative signal shapes (blue solid line) in different reconstructed $q^2$ bins.
		The contributions of different signal and various background components are indicated, where other background includes all other components.}
\end{figure*}

Figure~\ref{fig:fit6} shows the results of the fits to the $M_{\pi^+\pi^-}$ distributions in different reconstructed $q^2$ bins,
in which the shapes the $D^+\to f_0(500)\mu^+\nu_\mu$ and $D^+\to f_0(500)e^+\nu_e$ signals are described with the alternative
RBW shapes.

\begin{figure*}[htbp]
	\centering
	\includegraphics[width=0.9\linewidth]{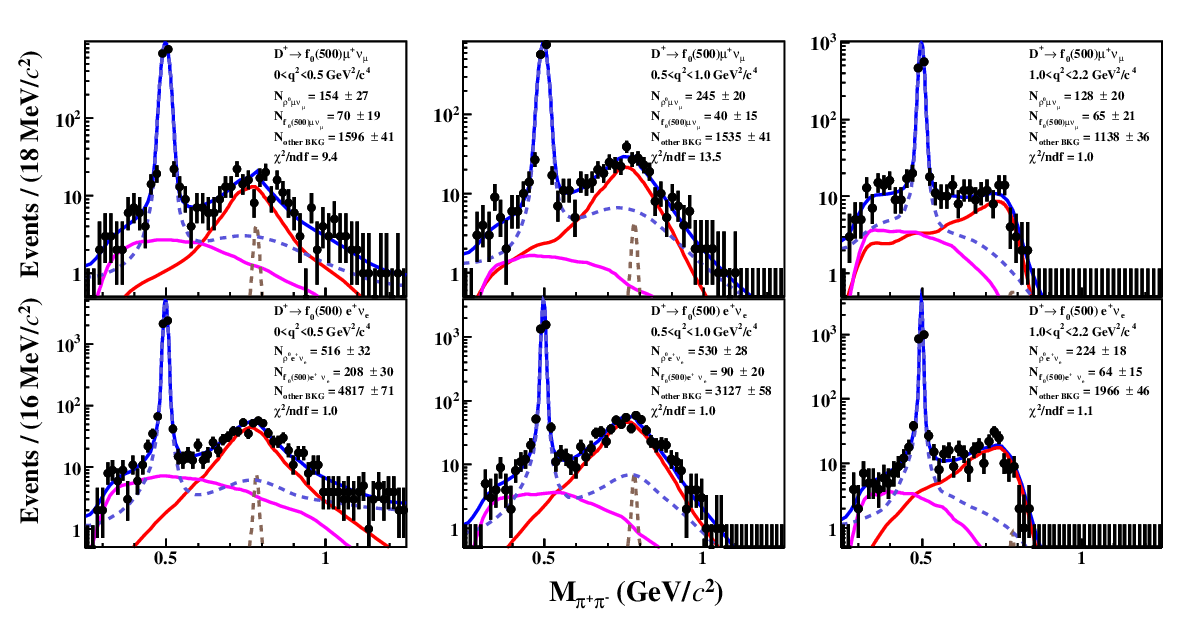}
	\caption{\label{fig:fit7}
			Fits to the $M_{\pi^+\pi^-}$ distributions of the candidates for
		(up) $D^+\to \pi^+\pi^-\mu^{+}\nu_{\mu}$ and (down) $D^+\to \pi^+\pi^-e^{+}\nu_e$ in different reconstructed $q^2$ bins,
		in which the parameter of other background shape is increased  by 0.5.
		The contributions of the signal and various background components are indicated, where other background includes all other components.}
\end{figure*}

\begin{figure*}[htbp]
	\centering
	\includegraphics[width=0.9\linewidth]{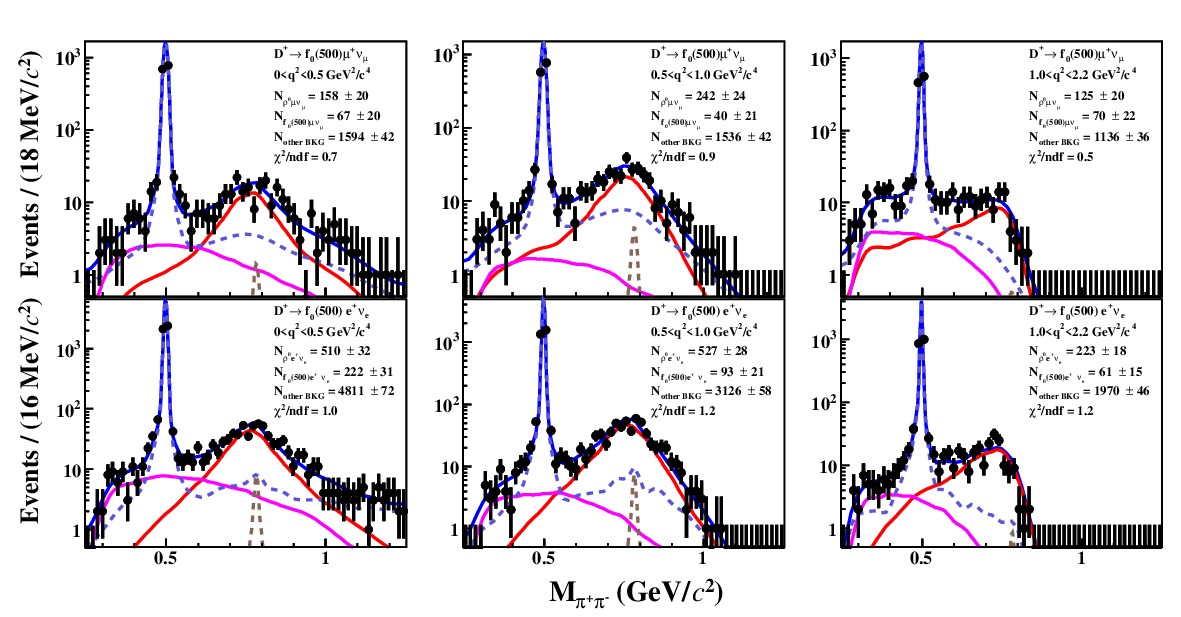}
	\caption{\label{fig:fit8}
	Fits to the $M_{\pi^+\pi^-}$ distributions of the candidates for
(up) $D^+\to \pi^+\pi^-\mu^{+}\nu_{\mu}$ and (down) $D^+\to \pi^+\pi^-e^{+}\nu_e$ in different reconstructed $q^2$ bins,
in which the parameter of other background shape is decreased  by 0.5.
The contributions of the signal and various background components are indicated, where other background includes all other components.}
\end{figure*}

Figures~\ref{fig:fit7} and \ref{fig:fit8} show the fits to the $M_{\pi^+\pi^-}$ distributions of $D^+\to f_0(500)\mu^+\nu_\mu$ in different reconstructed $q^2$ bins,
in which the parameters of other background shapes are increased and decreased by 0.5, respectively.

\end{document}